\documentclass[runningheads]{llncs}
\usepackage{graphicx}
\usepackage[T1]{fontenc}
\usepackage{textcomp,gensymb}
\usepackage{amsmath}
\usepackage{amsfonts}
\usepackage{array}

\begin{document}

\title{Non-invasive Assessment of Hepatic Venous Pressure Gradient (HVPG) Based on MR Flow Imaging and Computational Fluid Dynamics}
\titlerunning{MRI-based Hepatic Venous Pressure Gradient}


\author{Kexin Wang % 1{Wang, Kexin}
\inst{1}
\and
Shuo Wang % 2{Wang, Shuo}
\inst{2,3}
\and
Minghua Xiong % 3{Xiong, Minghua}
\inst{4}
\and
\\
Chengyan Wang % 4{Wang, Chengyan}
\inst{5}
\and
He Wang % 5{Wang, He}
\inst{5,6,7}}
\authorrunning{K. Wang et al.}
\institute{Department of Physics, Fudan University, Shanghai, China 
\and
Digital Medical Research Center, School of Basic Medical Sciences, Fudan University, Shanghai, China
\and
Shanghai Key Laboratory of Medical Image Computing and Computer Assisted Intervention, Shanghai, China
\and
Shanghai Zhiyu Software Information Co., Ltd
\and
Human Phenome Institute, Fudan University, Shanghai, China
\and
Institute of Science and Technology for Brain-inspired Intelligence, Fudan University, Shanghai, China
\and
Key Laboratory of Computational Neuroscience and Brain-Inspired Intelligence (Fudan University), Ministry of Education, China \\
\email{\{wangcy,hewang\}@fudan.edu.cn}\\
\url{http://homepage.fudan.edu.cn/hewang} 
}
\maketitle              % typeset the header of the contribution

\begin{abstract}
Clinically significant portal hypertension (CSPH) is a severe complication of chronic liver disease associated with cirrhosis, which is diagnosed by the measurement of hepatic venous pressure gradient (HVPG). However, HVPG measurement is invasive and therefore difficult to be widely applied in clinical routines. There is no currently available technique to measure HVPG noninvasively. Computational fluid dynamics (CFD) has been used for noninvasive measurement of vascular pressure gradient in the intracranial and coronary arteries. However, it has been scarcely employed in the hepatic vessel system due to the difficulties in reconstructing precise vascular anatomies and setting appropriate boundary conditions. Several computer tomography and ultrasound based studies have verified the effectiveness of virtual HVPG (vHVPG) by directly connecting the portal veins and hepatic veins before CFD simulations \cite{CHESS1601,CHESS1701}. We apply the latest techniques of phase-contrast magnetic resonance imaging (PC-MRI) and DIXON to obtain the velocity and vessel anatomies at the same time. Besides, we improve the CFD pipeline in regards to the construction of vessel connections and reduction of calculation time. The proposed method shows high accuracy in the CSPH diagnosis in a study containing ten healthy volunteers and five patients. The MRI-based noninvasive HVPG measurement is promising in the clinical application of CSPH diagnosis.

\end{abstract}
\section{Introduction}
Portal hypertension is a severe complication among patients with chronic liver disease. Currently, clinically significant portal hypertension (CSPH) is diagnosed with the measurement of hepatic venous pressure gradient (HVPG) larger than 10 mmHg. This HVPG measurement is conducted by putting a balloon catheter into the right hepatic vein to get both the free hepatic venous pressure (FHVP) and the wedged hepatic venous pressure (WHVP). The HVPG is defined as the difference between WHVP and FHVP. Although the gold standard of clinical diagnosis of CSPH is HVPG, it is invasive, probably leading to infection\cite{Fang2020Consensus,Franchis2016Portal} and other complications \cite{2018Liver}, thus inaccessible to some patients. 

Recently, several non-invasive measurements have been proposed to obtain HVPG, transient elastography (TE) for example\cite{Tana2018Diagnosing,Cast2009Early}. However, TE is often unreliable for patients with obesity or intrahepatic inflammatory activities \cite{De2015Expanding}. Some serum biomarkers including prothrombin index (PI) and aspartate aminotransferase (AST) to alanine aminotransferase (ALT) ratio (AAR) are more widely available in hospitals, but their accuracy is lower than TE. 

Computational fluid dynamics (CFD) simulation has grown rapidly in the past few years for the diagnosis of cardiovascular diseases\cite{MICCAI1,MICCAI2,MICCAI3}, carotid artery diseases\cite{Jenkins2003Noninvasive} and intracranial artery occlusive diseases\cite{2008Large}. The virtual blood pressure gradient is measured by solving the Navier-Stokes equation with certain boundary conditions. CFD-based measurement of blood pressure gradient has achieved remarkable success in assessing intracranial and coronary arteries (e.g., cardiac fractional flow reserve (FFR)), while it is less explored in the hepatic vascular system due to two possible reasons. First, accurate CFD simulation is highly dependant on the precise reconstruction of vascular anatomy and the measurement of inflow velocity. However, the anatomies of hepatic vein and portal vein are difficult to be obtained. Currently available ultrasound (US) based velocity measurement is not robust and repeatable. Second, the computational efficiency is another limitation for its application in clinical routines due to the complex structure of liver vessels. 

This paper proposes a one-stop solution for the noninvasive measurement of HVPG in both the healthy volunteers and patients with liver cirrhosis by using multi-contrast magnetic resonance imaging (MRI) and an efficient CFD model. 

\section{Methods and Experiments}
\subsection{Multi-contrast MRI}
To avoid the use of contrast agents, multi-echo DIXON (mDIXON) imaging is conducted to obtain the anatomy of hepatic vessels in this study. A 3D gradient-echo (GRE) sequence with multiple echoes is used. All the subjects are scanned on a 3.0 T MR scanner (Ingenia, Philips, the Netherlands). The imaging parameters are: repetition time (TR) = 3.4 ms; echo time (TE) = 0.95 ms; field-of-view (FOV) = 350 × 350 × 120 mm$^3$; flip angle (FA) = 10°; spatial resolution = 0.87 × 0.87 × 2.0 mm$^3$, which sets the limit for the smallest vessel reconstructed; SENSitivity Encoding (SENSE) factor = 2; scan time = 12.5 ms.

Blood flow velocity is measured using phase-contrast (PC) GRE sequence. Imaging is performed during free breathing with cardiac and respiratory gatings. The imaging parameters for PC-MRI are: TR = 2.9 ms; TE = 1.45 ms; FOV = 300 × 300 × 300 mm$^3$; FA = 6°; spatial resolution = 0.87 × 0.87 × 2.0 mm$^3$; SENSE factor = 2; encoding velocity = 40 cm/s. The acquisition time is nearly 2.5 minutes depending on the subject’s heart rate. The blood flow acquisition plane is selected perpendicular to the portal vein on the morphological scout sequences.

\subsection{Image Analysis}
MRI data are transferred in Digital Imaging and Communications in Medicine (DICOM) format to a workstation for analysis. The vessels are automatically extracted from the mDIXON images using an pre-trained deep neural networks (DNN)\cite{Zhang2020Robust}. Two separate pre-trained deep neural networks (DNNs) are applied to obtain the portal veins and hepatic veins automatically. The DNNs are trained on an additional dataset including 30 manually labelled cases in our previous work.

Extraction of the temporal flow velocity curve during a cardiac cycle is from the cross section of portal vein manually labeled by an experienced radiologist on the PC-MRI images. The flow velocity is calculated as the mean value of the cross section obtained during the 30 phases of several repeated cardiac cycles (see the process in Fig. 1. $\boldsymbol{u}(t)$). 

\begin{figure}
\includegraphics[width=\textwidth]{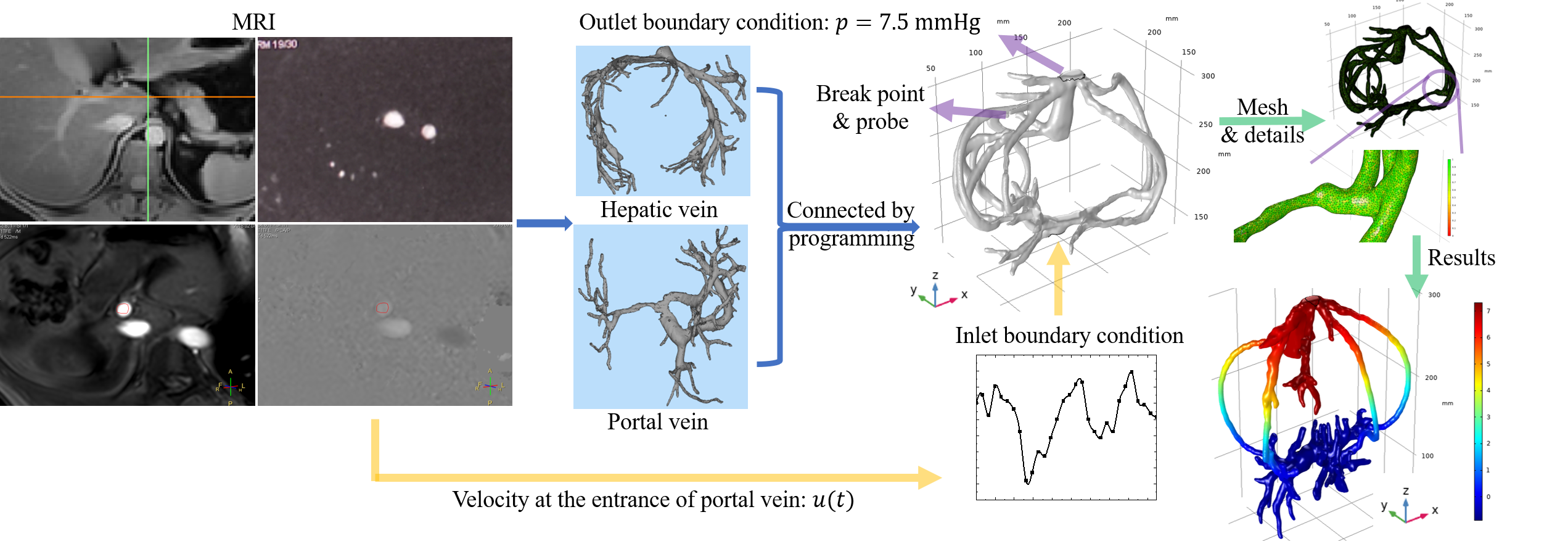}
\caption{The schematic diagram of the MRI based CFD model. From the MRI images on the left to the pressure distribution at the bottom right. Firstly, geometry of the hepatic vein and the portal vein are segmented individually, together with the attained velocity of flow at the entrance of the portal vein. The combination of the yellow and blue arrow is the green arrow, which represents applying the connected model and boundary conditions, then with meshing based on both practice, CFD calculation achieves the pressure distribution eventually.} 
\end{figure}
 
\subsection{CFD Simulation} 
CFD simulation can provide accurate pressure measurement based on the geometry of blood vessels and the boundary conditions, e.g., the velocity and the reference pressure at the inlet and the outlet of the blood vessels. Finite element method (FEM) is applied to solve the the Navier-Stokes (NS) equations. The main components of the simulation can be divided into three parts: a) dominant physics of the laminar fluid model, b) settings of the boundary conditions, and c) development of a solver in space-time.   
\subsubsection{Laminar Physics} 
A three-dimensional NS equations with the assumption of stationary incompressible fluid are applied in our study as the following:

\begin{equation}
\rho (\boldsymbol{u}\cdot\nabla)\boldsymbol{u}=\nabla\left( -p \mathbb{I} +\mu[\nabla\boldsymbol{u}+(\nabla\boldsymbol{u})^T]\right) +\boldsymbol{F}
\end{equation}

\begin{equation}
\rho\nabla\cdot\boldsymbol{u}=0
\end{equation}
where $\boldsymbol{u}$ is the flow velocity, $p$ is the pressure, $\mu$ is the dynamic viscosity and $\rho$ the density of the blood, which is assumed to be Newtonian fluid with $\mu$ = 0.005 Pa·s and $\rho$=1050 kg/m$^3$\cite{CHESS1601}. $\boldsymbol{F}$ is the volume force vector and is set to be zero as we ignore the gravity of blood. The Eq.1 shows the balance of momentum from Newton's Second Law, and the Eq.2 represents the constraint of continuity, which is incompressible. 

We use only laminar physics here rather than the turbulence and non-Newtonian modelling, or Fluid-Structure Interaction (FSI) model, due to the small Reynolds number (Re = 27) and our consideration of vascular sclerosis. On the other hand, although the geometry of vessels counts a lot in the CFD model and admittedly, it will be more accurate if we use FSI, patients suffered from CSPH experience significantly reduced wall compliance, rendering our assumption of rigid wall reasonable and computationally efficient.

\subsubsection{Boundary Condition}
The inlet is the entrance of the portal vein, with the velocity set as a continuous function of time by interpolating cubic spline function into 30 uniformly-spaced time intervals in a cardiac cycle. Although the velocity has already been averaged over time, we prolong the simulation time of interest by duplicating the cardiac cycle four times (see in fig. 2 (b)), guaranteeing that the flow is fully developed. A smooth transition from zero is added at the beginning 100 ms to accelerate the convergence, as enlarged in fig. 2 (a). 

The outlet is naturally the exit of the hepatic vein, and a reference pressure is set here. As vHVPG is calculated as the pressure difference before and after the balloon is inflated, it is irrelevant to the absolute value of the reference pressure. For the sake of similarity to clinical measurement and following the physiological condition, the constant value of 7.5 mmHg is set as the outlet pressure condition.
 
\begin{figure}
\includegraphics[width=\textwidth]{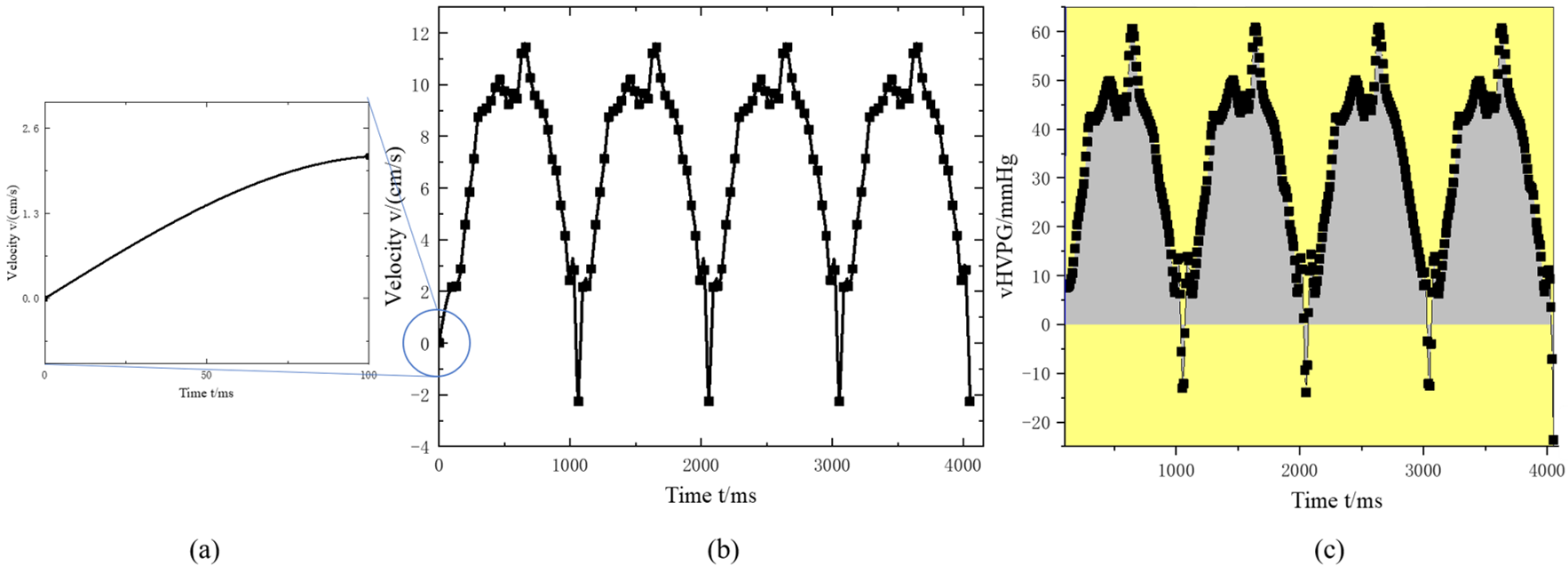}
\caption{Velocity settings in an example of a patient with liver cirrhosis. (a) The illustration of how to smooth the abrupt transition at the beginning 100 ms of a flow velocity period. (b) The exact interpolated function of velocity consists of four duplicate periods. (c) Corresponding vHVPG as a function of time. The grey denotes the section enclosed by the function and the x-axis, which is the integration of vHVPG during the calculation time considering the plus and minus sign. Then the integration by the time is the average vHVPG.} 
\end{figure}

\subsubsection{Solver} 

We use the Laminar Fluid module of COMSOL Multiphysics for finite element simulation. We use a fully coupled direct solver for CFD models. To avoid run-out-of-memory error, we cut the iteration to be ten thousand times. Two dependent variables are coupled in the equation, i.e., the pressure and velocity field. Moreover, algebraic multigrid (AMG) solvers are applied to obtain these two values at each node\cite{Falgout2006An}. 

\subsection{Experiments}
This retrospective study includes a total of ten healthy volunteers and five patients with liver cirrhosis. The study was approved by the ethics committee of the local hospital. Informed consent was obtained from all the subjects prior to the examinations. The cirrhosis was diagnosis by means of liver biopsy. 

\subsubsection{Building Geometry}
Portal veins and hepatic veins are extracted from PC-MRI separately and therefore the connection of capillary is needed. Since CFD largely depends on the geometry, the strategy to rebuild it is fundamental and original here in our work. As can be viewed in Fig.3, in principle, the connection of portal vein and hepatic vein should get close to the reality, where both one-to-one and one-to-many connections exist. Therefore, the decisions to carry out which kind of the link and how to bridge the neighbour ports are important, and we demonstrate our procedure as below.

\begin{figure}
\includegraphics[width=\textwidth]{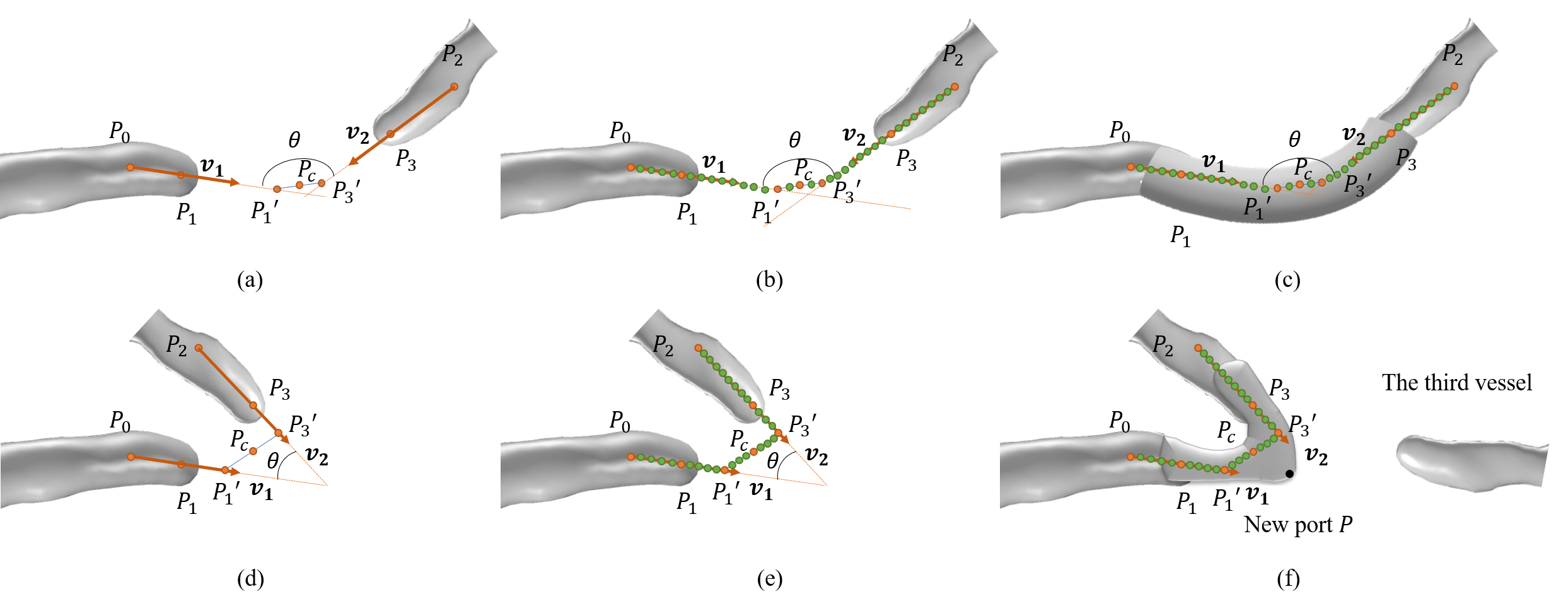}
\caption{Procedure to connect vessels by interpolation. The top line is the case when the angle between two vessels $\theta>90\degree$, and the bottom line is $\theta<90\degree$. $\protect\overrightarrow{P_0P_1}$ defines the directional vector $\boldsymbol{v_1}$ of the vessel on the left and $\protect\overrightarrow{P_2P_3}$ works for $\boldsymbol{v_2}$ on the right. $P_1'$ (or $P_3'$) is the translation of $P_1$ (or $P_3$) along $\boldsymbol{v_1}$ (or $\boldsymbol{v_2}$) at the distance of $|P_1P_3|/2$. The midpoint $P_c$ is in the middle of $P_1'$ and $P_3'$. In (b) and (e), interpolation is demonstrated as green dots along $P_0, P_1, P_1', P_c, P_3', P_3$. As in (c), large $\theta$ results in a smooth bridge and no more connections are allowed for them, while in (f) the curve will provide a new point $P$, served as a new port to be connected with a third vessel.}
\end{figure}

First of all, find two ends of the axis in two vessels, say ${P_0, P_1, P_2, P_3}$, from which the direction vector $\boldsymbol{v_1}$ and $\boldsymbol{v_2}$ will be obtained. Then defining the center point $P_c$ by the midpoint of $P_1'P_3'$, where the spot $P_1'$ is translated from $P_1$ along the vector $\boldsymbol{v_1}$ by the distance of $|P_1P_3|/2$ and the same for $P_3'$ but from $P_3$ along $\boldsymbol{v_2}$. Next, interpolate enough points between the point set of $P_0, P_1, P_1', P_c, P_3', P_3$ and extrude the circle surface with the diameter of initial diameter on both sides along the interpolated curve. Finally, with some polishing, the connection is built up. However, if the angle between two vectors $\boldsymbol{v_1}$ and $\boldsymbol{v_2}$ is acute, the bridge will end up with a rather sharp curve, generating a new point $P$ to be connected with other vessels. 

\begin{figure}
\includegraphics[width=\textwidth]{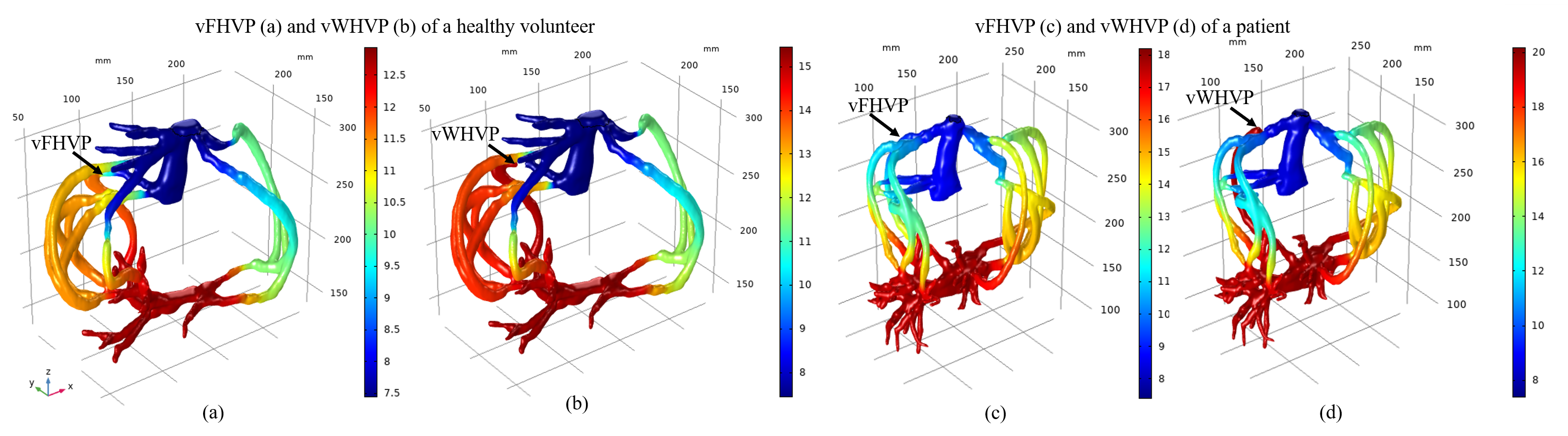}
\caption{Average pressure distribution of typical CFD model. The first two pictures show the pseudo-color images of flow pressure for a healthy volunteer and the other two are for a patient with liver cirrhosis. Arrows show measurement location. As for virtual free hepatic venous pressure (vFHVP), (a)=9.8 mmHg, (c)=9.6 mmHg; for virtual wedged hepatic venous pressure (vWHVP), (b)=14.5 mmHg, (d)=19.8 mmHg. vHVPG = vWHVP - vFHVP, and it is 4.7 mmHg for the healthy person, but 10.2 mmHg for the patient with liver disease. As CSPH is classified as HVPG $\geq$ 10 mmHg, the findings of these two cases fulfill this standard, therefore verifying our method of CFD model based on PC-MRI.} 
\end{figure}

After the connection, a break point is created for the modeling of WHVP, representing a strict block of the blood flow along the vessel. We ring cut the right-most secondary branch at the point of about 1 cm away from the fork. Afterwards, both sides are healed by the smooth smallest surface and then follows the remesh process. Remeshing includes uniforming the surface mesh and creating the volume mesh from the surface, providing an exact calculation domain for CFD with FEM. See in Fig. 4, pressure is evaluated near the break point from the side closer to the portal vein. The virtual FHVP (vFHVP) is from the model of complete blood vessels, while the virtual WHVP (vWHVP) is achieved in the cut-off model. Therefore, similar to the clinical measurement in practice, the virtual hepatic venous pressure gradient (vHVPG) is the difference between the above two pressures: vHVPG = vWHVP - vFHVP.    

\section{Results and Discussion}
In this section we present the simulation result and prove the accuracy by analyse the data. In fig. 4, typical examples of average pressure distribution in a healthy volunteer and a patient with liver cirrhosis can be compared with each other. Three main properties verify the CFD model primarily: 1) Flow pressure decreases gradually from the inlet at the portal vein to the outlet at the hepatic vein. 2) Averagely, vWHVP is larger than vFHVP in both cases. 3) Obviously, the flow pressure of the patient is larger than that of the healthy volunteer. The simulation result of vHVPG is concluded in Table 1. 

HVPG larger than 10 mmHg is defined as CSPH for the statistical analysis. Although the number of cases is limited, it still shows a significant increase of HVPG for the patients and 80 percents of them are found to be CSPH (HVPG $\leq$ 10 mmHg) according to our studies. Based on this confusing matrix, the kappa test has shown substantial consistency (k=0.7) between the diagnosis based on histopathology and our virtual HVPG measurements. Besides, the normal HVPG is suggested to be in the range of 3-5 mmHg for healthy adults\cite{De2015Expanding}, which is also fulfilled in our studies. 

\renewcommand{\arraystretch}{1.5}
\begin{table}

\caption{Summary of the vHVPG for Different Groups}\label{tab1}
\begin{tabular}{|l|l|l|l|l|l|l|}
\hline
{\bfseries Group}  & {\bfseries Average vHVPG (mmHg)} & {\bfseries No. of true} & {\bfseries No. of false}\\
\hline
Healthy volunteers & $5.9\pm 2.3$ & 9 ($\leq 10$mmHg) & 1 ($\geq 10$mmHg)\\
Patients with liver cirrhosis &  $21.5\pm 8.8$ & 4 ($\geq 10$mmHg) & 1 ($\leq 10$mmHg)\\
\hline
\end{tabular}
\end{table}

As for the next step, the interaction between structures and the fluid will be considered, resulting in a more precise model than the current one of simple single-phase laminar physics. Also, in order to virtually model the the invasive measurement as close as possible, the geometry ought to be improved as well. Although the spatial resolution of CT is usually higher than MRI for liver imaging, the high-resolution 3D GRE sequence used in this study achieves resolution of 0.87 × 0.87 × 2.0 mm$^3$. We believe that the resolution is comparable to conventional CT and adequate for CFD modelling. Interestingly, although the fine-scale capillaries are not modelled, the vHVPG result is consistent with histopathology diagnosis. It has been reported that the side branches do affect some specific wall variables of CFD simulation results, e.g. wall shear stress, although it is in the cardiovascular system \cite{add1,add2,add3}. However, it was suggested that including side blood vessels up to 1 mm in diameter is appropriate \cite{add3}. While the vessel diameter that can be observed in our study is no less than 0.87 mm corresponding to the spatial resolution, we believe such simplification has a limited influence on the calculation of pressure gradient. However, we still believe that the way forward lies in some modified CFD model including nearly all the vessels or at least their effects, constructed automatically with as little manual interference as possible.   

\section{Conclusion}
In this article, we introduced a one-stop solution to acquire an accurate estimate of HVPG non-invasively, assisted by PC-MRI and CFD. For the first time, CFD model is established entirely based on MRI for the geometry and boundary conditions of velocity in the liver system, as we take advantage of the latest technology of PC-MRI. Since the limit of resolution forbids a complete map of entire blood vessels in the liver, trunks of hepatic vein and portal vein are preserved and connected automatically by our original programming method, which demonstrates complex networking very similar to the real vessels. FEM solver is applied in the CFD model accelerating the calculation speed. By now, it only takes less than twenty minutes to compute HVPG from the raw image, while it can even be faster in our further study. Testing in a data set of ten healthy people and five patients with liver cirrhosis, we verify the accuracy of our method in diagnosing people with CSPH. Our study facilitates the process of non-invasive and patient-specific treatment.         

\subsubsection{Acknowledgements} This study was funded by the National Natural Science Foundation of China (No. 81971583, No. 62001120), National Key R \& D Program of China (No. 2018YFC1312900, No. 2019YFA0709502), Shanghai Natural Science Foundation (No. 20ZR1406400), Shanghai Municipal Science and Technology Major Project (No.2017SHZDZX01, No.2018SHZDZX01), Shanghai Municipal Science and Technology (No.17411953600), Shanghai Sailing Program (No. 20YF1402400), ZJLab and Key Laboratory of Computational Neuroscience and BrainInspired Intelligence (Fudan University), Ministry of Education, China.


\bibliographystyle{splncs04}
\bibliography{ref}

\end{document}